**Comment for article "Limits on a nucleon-nucleon monopole-dipole (axionlike) P,T-noninvariant interaction from spin relaxation of polarized 3He" Yu. N. Pokotilovski**

A.P. Serebrov
*Petersburg Nuclear Physics Institute RAS*

**Abstract**

In article "Limits on a nucleon-nucleon monopole-dipole (axionlike) P,T-noninvariant interaction from spin relaxation of polarized 3He" Yu. N. Pokotilovski arXiv:0902.1682v2 [nucl-ex] 12 Feb 2009 it was presented new constraints for the product of the scalar, pseudo-scalar dimensionless constants: $g_S g_P < 1 \cdot 10^{-17}$ for the parameter $\lambda$ about $10^{-4}$ cm, determining the range of forces. In this comment it is shown that the real constraint for the same parameter $\lambda$ from the same experimental data is about $2 \cdot 10^{-11}$. It is worse in about one hundred times that the constraint from UCN depolarization and can not be considered as a new limit on a nucleon-nucleon monopole-dipole (axionlike) P,T-noninvariant interaction.

It would be interesting to compare the sensitivity of experiments of UCN depolarization and $^3$He depolarization for estimation limits on $g_S g_P$ constraints. These estimations were done in work [1] for UCN depolarization and in work [2] for $^3$He depolarization. It seems for the first view that experiments with $^3$He depolarization could be more sensitive because the time of relaxation of $^3$He depolarization is about a few hundred hours. For example, the spin relaxation time ($T_{^3\mathrm{He}}$) was observed equal to 662 h for cell with size 5 cm in the guiding magnetic field 10 G and at the pressure of $^3$He 0.97 atm [3]. Whereas the spin relaxation time for UCN is about a few hundred second [1]. Nevertheless it will be shown below that sensitivity of experiments with $^3$He depolarization is less than with UCN depolarization.

It is clear that the frequency of collisions of $^3$He atom with cell walls is considerably suppressed at pressure of about 1 atm due to collisions between $^3$He atoms. The length of diffusion of $^3$He atom is proportional free path and root squared from number of collision. Therefore $a \approx \sqrt{n}\lambda_{\mathrm{He,He}}$, here $a$ is size of cell, $\lambda_{\mathrm{He,He}}$ is free path of $^3$He atom between collisions, $n$ is the required number of collisions for diffusion between walls. Thus the frequency of collisions of $^3$He atom with walls ($f_{\mathrm{wall}}$) is:

$$f_{\text{wall}} = f_{\text{He,He}} / n = f_{\text{He,He}} \left( \frac{\lambda_{\text{He,He}}}{a} \right)^2 = 0.27 \text{ s}^{-1}, \tag{1}$$

where the frequency of collisions of $^3$He atoms ($f_{\text{He,He}}$) at the pressure about 1 atm is about $3.3 \cdot 10^9$ s$^{-1}$, $\lambda_{\text{He,He}} = 4.5 \cdot 10^{-5}$ cm, $a = 5$ cm. We can estimate the probability of $^3$He atom depolarization per one collision with cell walls:

$$\beta_{^3\text{He}} = \frac{1}{T_{^3\text{He}} \cdot f_{\text{wall}}} = 1.5 \cdot 10^{-6}. \tag{2}$$

Now we can estimate $g_S g_P$ constraints, for example at $\lambda = 10^{-4}$ cm (we chose the range of CP-violating forces $\lambda = 10^{-4}$ cm because it is just about free path of $^3$He atom near the wall). We can use formulas from work [1, 4] to obtain the final formula for estimation of $g_S g_P$:

$$g_S g_P \leq \left. \frac{\sqrt{2\beta_{^3\text{He}}} \, m_n v_{^3\text{He}}}{\hbar \lambda^2 N} \right|_{\lambda = 10^{-4} \text{ cm}} = 2 \cdot 10^{-11}. \tag{12}$$

Where $N$ is nucleon number density $N$, $m_n$ is the neutron mass, $v_{^3\text{He}}$ is the component of the $^3$He atom velocity normal to a wall surface. We can see [1] that this limit is about two orders of magnitude worse than limit from UCN depolarization.

Here we have to mention that our estimation for the $g_S g_P (\lambda = 10^{-4}$ cm) from $^3$He depolarization is bigger about $10^6 \div 10^7$ orders of magnitude, than in the work [2]. The result of work [2] is not correct because the estimation of $g_S g_P$ limit was done using the formula of adiabatic conditions instead of formula for non-adiabatic conditions. The more detailed explanation of this mistake was discussed in [4]. The critical remarks of work [4] were accepted by author in the third version [5] of article for estimation of $g_S g_P$ from UCN depolarization. But estimations for $g_S g_P$ from $^3$He depolarization are uncorrected yet. Therefore we are forced to mention about this situation. Moreover it is very important question (what is the real $g_S g_P$ limit for $\lambda$ about $10^{-4}$ cm) for the planning of future experiments.

**References**

[1] A.P. Serebrov, arXiv:0902.1056v1., Physics Letters B 680 (2009) 423–427

[2] Yu.N. Pokotilovski, arXiv:0902.1682v2.

[3] R. Parnell et al., Nucl. Instr. Meth. A **589** (2009) 774.

[4] A.P. Serebrov, arXiv:0905.4137v1.

[5] V.K.Ignatovich, Yu.N.Pokotilovski, arXiv:0902.3425v3